\documentstyle[amssymb,aps]{revtex}
\begin{document}
\draft
\title{Sub-Poissonian statistics in order-to-chaos transition}
\author{Gagik Yu. Kryuchkyan$^{1,2}$ and Suren B. Manvelyan$^{2}$}
\address{$^{1}$Yerevan State University, Manookyan 1, Yerevan, 375049, Armenia\\
and\\
$^{2}$Institute for Physical Research, National Academy of Sciences,\\
Ashtarak-2, 378410, Armenia}
\maketitle

\begin{abstract}
We study the phenomena at the overlap of quantum chaos and nonclassical
statistics for the time-dependent model of nonlinear oscillator. It is shown
in the framework of Mandel Q-parameter and Wigner function that the
statistics of oscillatory excitation number is drastically changed in
order-to chaos transition. The essential improvement of sub-Poissonian
statistics in comparison with an analogous one for the standard model of
driven anharmonic oscillator is observed for the regular operational regime.
It is shown that in the chaotic regime the system exhibits the range of sub-
and super-Poissonian statistics which alternate one to other depending on
time intervals. Unusual dependence of the variance of oscillatory number on
the external noise level for the chaotic dynamics is observed. The scaling
invariance of the quantum statistics is demonstrated and its relation to
dissipation and decoherence is studied.
\end{abstract}

\section{Introduction}

The subject of this article is at the boundary of two basic phenomena in
quantum physics nowadays attracting significant interest of a broad
readership. One of them is quantum chaos{\bf , }which has been intensively
investigated in last two decades and has been found in many nonlinear
systems. Quantum effects inherent to the chaotic behavior of classical
systems are an interesting field pertaining to many problems of fundamental
interest \cite{Casati}. Another phenomenon is the generation of nonclassical
states, including nonclassical states of light with a range of unique
properties, e.g. sub-Poissonian statistics and squeezing. At present the
problem of nonclassical states generation draws a lot of attention, both
from the standpoints of pure knowledge and possible applications \cite
{Davidovich}. The principal purpose of the present paper is to study the
interesting crossovers between these phenomena for an open time-dependent
quantum system.

The majority of studies of quantum chaos for isolated or so called
Hamiltonian systems, the classical counterparts of which are chaotic, focus
on static properties such{\bf \ }as spectral statistics of energy levels and
transition probabilities between eigenstates of the system. A variety of
studies have also been carried out to understand{\bf \ }the features of
time-dependent chaotic systems. By contrast to that, very little work has
been done to investigate the quantum chaos for open nonlinear systems. The
beginning of study of an open chaotic system can be dated back to the papers 
\cite{Graham}, where the authors have analyzed the kicked rotor and similar
systems with discrete time interacting with a heat bath. Quite generally,
chaos in classical conservative and dissipative systems with noise has
completely different properties, e.g., strange attractors can appear only in
dissipative systems. Noise can induce a number of interesting phenomena in
transition from regular to chaotic behavior. In particular, it was shown
that under definite conditions noise can induce chaos as well as destroy
chaotic behavior (see for example \cite{Gao}). In recent years much effort
has been expended, both theoretically and experimentally, to explore the
role of quantum fluctuations and noise in order-to-chaos transition for open
systems. It is obvious, that the investigations in this area are connected
with the quantum-classical correspondence problem in general and with
environment induced decoherence and dissipation in particular. Recently this
topic has been the focus of theoretical investigations. As a part of these
studies it has been recognized \cite{Zurek1} that the decoherence has rather
unique properties for systems classical analogues of which are chaotic. In
particular, the formation of sub-Plank structure in phase space has been
discussed for chaotic system \cite{Zurek}. The connection between quantum
and classical treatments of chaos was also realized by means of comparison
between strange attractors on the classical Poincar\'{e} section and the
contour plots of the Wigner functions \cite{OurPRE}. Some studies have
explored the connection between quantum dynamical manifestation of chaos and
quantum entanglement which is possibly the most typical property of
composite quantum systems \cite{Entanglement}. From the experimental
viewpoint, observation of dissipative effects and environment induced
decoherence of dynamically localized states in the quantum delta-kicked
rotor is carried out with the gas of ultracold caesium atoms in a
magneto-optical trap subjected to a pulsed standing wave \cite{Amman},\cite
{Raizen99}. Recently, new problems of chaotic motion have been studied in
the experimental scheme with ultra-cold atoms in magneto-optical double-well
potential \cite{Deutsch}. Fluctuation and decoherence in three-level model
of chaos-assisted tunneling have been experimentally studied for samples of
caesium atoms in an amplitude-modulated standing wave of light \cite
{Raizen02}.

The other problem of interest in this area relates to the production of
various nonclassical states in systems with chaotic dynamics. Generation of
non-classical states of light has been the subject of an intense theoretical
and experimental activity since the first observations of squeezed states of
light. Generation of squeezed states in chaotic systems was studied
numerically and analytically by using a specific form of perturbation theory
for the model of quantum kicked rotator in Refs. \cite{Alekseev}. The result
of numerical experiments on quadrature squeezing in simple quantum models
that allow the transition to chaos in the classical limit are presented in 
\cite{Xie}. In a recent paper sub-Poissonian statistics of oscillatory
excitations numbers is established for chaotic dynamics of nonlinear
oscillator \cite{OurPRL}. The study of these phenomena provides a
fundamental understanding of quantum fluctuations in quantum chaos and opens
a way for new experimental studies of the quantum dissipative chaos in the
field of Quantum Optics.

In addition to these important developments in investigations of quantum
fluctuations and noise assisted to chaotic motion, in the present paper we
examine mainly quantum statistics of elementary excitations for the model of
nonlinear dissipative oscillator which shows the order-to-chaos transition.
Our goal is twofold: first we study in detail what \ kind of statistics of
oscillatory number states takes place in the quantum chaos, for both cases
of vacuum- and temperature-reservoir, and second we analyze Wigner functions
in a chaotic regime and strange attractors from the point of view of
nonclassical statistics. It should be noted that sub-Poissonian statistics
of oscillatory excitation numbers was established for chaotic dissipative
dynamics and for the case of vacuum-reservoir in our previous paper \cite
{OurPRL}. In the part of the present paper we essentially expand this study
particularly considering the important details as well as the case of
coupling of the nonlinear oscillator with nonzero temperature reservoir. The
other part of the paper is devoted to the problems of generation of
sub-Poissonian light, parameter scaling in chaotic dynamics and to the
external noise-induced effects.

The requirement in realization of this study is to have a proper quantum
model showing both regular and chaotic dynamics in the classical limit. We
propose a nonlinear oscillator driven by two forces at different frequencies
for this goal. This model was proposed to study the quantum chaos in one of
the our previous papers \cite{OurPRE}, where it was shown that the model is
apt to verification in experiments. This model also is attractive for the
following reasons: (i) it is different from a single driven nonlinear
oscillator and kicked rotor, where a pulsed pump field could be used for
realization of a chaotic regime, and may be proposed for experimental
studies of quantum chaos in the area of quantum optics with a cw laser; (ii)
dynamics of this model exhibits a rich phase-space structure including
regimes of regular and chaotic motion depending on the system's parameters \
as well as displays remarkable quantum properties, in particular -
nonclassical sub-Poissonian statistics of elementary excitations.

Open quantum systems are usually studied in the framework of reduced density
matrix obtained by tracing over the degrees of freedom of environment. There
have been suggestions (see, for example,\cite{Spiller}) that chaotic
dissipative dynamics should be described on individual quantum trajectories
in the framework of quantum state diffusion method (QSD) \cite{Gisin}. In
addition to these studies, here we describe the quantum dissipative chaos
using a statistical ensemble of trajectories which is usually realized in
nature. Such description allows us to consider quantum manifestation of
chaos in measurable quantities.

The plan of this paper is as follows. In Sec.II nonlinear oscillator driven
by two forces at different frequencies is described and relevant regular and
chaotic operational regimes are discussed. In Sec. III the formation of
sub-Poissonian statistics of oscillatory excitations number for regular
dynamics is analyzed and application to the generation of strong
sub-Poissonian light is considered. The important novelty here is that the
radial squeezing effect is much stronger, than an analogous one for the
model of a single driven anharmonic oscillator. In Sec. IV we identify the\
kind of statistics of oscillatory number states which takes place for
quantum chaos. Our central result here is that nonclassical, sub-Poissonian
statistics can be realized for chaotic dynamics of the system under
consideration. We also discuss the role of the dissipation in scaling
relationships for the regular and chaotic motion as well as the dependence
of results on the{\bf \ }level of external noise. We summarize our results
in Sec. V.

\section{Order to chaos transition in doubly driven nonlinear oscillator}

In this section we briefly review the results on the model \cite{OurPRE}. We
treat dissipation and decoherence microscopically using a master equation
which is solved numerically in the framework of QSD approach. This model is
a nonlinear oscillator driven by two periodic monochromatic forces at
different frequencies $\omega _{1}$ and $\omega _{2}$ for which the
evolution of the reduced density operator $\rho $ is governed by the master
equation

\begin{equation}
\frac{\partial \rho }{\partial t}=-\frac{i}{\hbar }\left[ H_{0}+H_{int,}\rho %
\right] +\sum_{i=1,2}\left( L_{i}\rho L_{i}^{+}-\frac{1}{2}%
L_{i}^{+}L_{i}\rho -\frac{1}{2}\rho L_{i}^{+}L_{i}\right) .  \label{mastereq}
\end{equation}
We use the interaction picture corresponding to the transformation $\rho
\rightarrow e^{-i\omega _{1}a^{+}at}\rho e^{i\omega _{1}a^{+}at}$ and the
Hamiltonians are: 
\begin{eqnarray}
H_{0} &=&\hbar \Delta a^{+}a,  \label{hamiltonian} \\
H_{int} &=&\ \hbar \left[ \left( \Omega _{1}+\Omega _{2}\exp \left( -i\delta
t\right) \right) a^{+}+\left( \Omega _{1}^{\ast }+\Omega _{2}^{\ast }\exp
\left( i\delta t\right) \right) a\right] +\hbar \chi (a^{+}a)^{2}.  \nonumber
\end{eqnarray}
\ \ \ \ \ \ \ \ Here$\ \Delta =\omega _{0}-$ $\omega _{1}$\ is the detuning,
and $\delta =\omega _{2}-$ $\omega _{1}$ is the difference between driving
frequencies, which works as modulation frequency. $a,a^{+}$ are boson
annihilation and creation operators and $L_{i}$ are the Lindblad operators:

\begin{equation}
L_{1}=\sqrt{\left( N+1\right) \gamma }a,\;L_{2}=\sqrt{N\gamma }a^{+},
\label{Lindblad}
\end{equation}
where $\gamma $ is the spontaneous decay rate of the dissipation process,
and $N$ denotes the mean number of quanta of a heat bath. The couplings with
two driving forces are given by Rabi frequencies $\Omega _{1}$ and $\Omega
_{2}$, and $\chi $ is the strength of anharmonicity. The last terms in Eq. (%
\ref{mastereq}) concern the influence of the environment induced diffusion.
We have followed the standard approach \cite{Gardiner}, \cite{Weiss} to
dissipative quantum dynamics in the range of weak coupling of the{\bf \ }%
system with the reservoir under the condition: $\gamma \ll k_{B}T/\hbar ,$
where $k_{B}T$ is the Boltzman's constant times temperature. Equation (\ref
{mastereq}) is obtained in both the rotating wave and Markov approximations,
without regard to{\bf \ }the driving-induced noise effects \cite{Kohler}.

To study the pure quantum effects we focus on the cases of very low
reservoir temperatures which, however, ought to be still larger than the
characteristic temperature $T\gg T_{ch}=\hbar \gamma /k_{B}.$ This
restriction implies that dissipative effects can be described
selfconsistently in the frame of the Lindblad equation (\ref{mastereq}).
Note here that for $N\ll 1$ and, even for vacuum reservoir, for $N\simeq 0$
the above restriction is valid for the majority of problems of quantum
optics.

For $\Omega _{2}=0$ this equation describes the single driven, dissipative
anharmonic oscillator, which is a well-known and archetypal model in
nonlinear physics \cite{Drummond}. In case of doubly driven oscillator ($%
\Omega _{2}\neq 0$), the interaction Hamiltonian (\ref{hamiltonian}) is
explicitly time-dependent and the system exhibits regions of regular and
chaotic motion. It should be noted that this model corresponds to a modified
model of Duffing oscillator, i.e. an oscillator where the oscillatory
frequency depends on the amplitude. To illustrate it, we first{\bf \ }pay
attention to the classical description.

\subsection{Classical equation and connection to the Duffing oscillator}

In the classical limit, the corresponding equation of motion for the
dimensionless amplitude $\alpha (t)=\left\langle a(t)\right\rangle =Tr(a\rho
(t))$ has the form 
\begin{equation}
\frac{d}{dt}\alpha =-\frac{1}{2}\gamma \alpha -i\left( \Delta +\chi
(1+2\left| \alpha \right| ^{2})\right) \alpha -i\left( \Omega _{1}+\Omega
_{2}\exp \left( -i\delta t\right) \right) .  \label{clas}
\end{equation}

It is easy to demonstrate that this equation is the rotating-wave
approximation of the equation for Duffing oscillator driven by two periodic
forces. Indeed, let us start with the following equation for a
time-dependent real amplitude

\begin{equation}
\ \stackrel{\cdot \cdot }{E}+\gamma \stackrel{\cdot }{E}+\omega _{0}^{2}[1+%
\frac{2\chi }{\omega _{0}}(1+E^{2}/2)]E=4\omega _{0}(\Omega _{1}\cos \omega
_{1}t+\Omega _{2}\cos \omega _{2}t).  \label{E}
\end{equation}
$\ $which describes such oscillator. This equation is a generalization of
the well-known Duffing oscillator to the case of two driving forces at
different frequencies $\omega _{1}$ and $\omega _{2}$. To perform the
rotating-wave approximation of the equation (\ref{E}) we introduce the
dimensionless complex amplitude $\alpha (t)$ as

\begin{equation}
E(t)=\alpha (t)\exp (-i\omega _{1}t)+c.c.  \label{E1}
\end{equation}

Neglecting also the second-order derivative of the slowly varying amplitude $%
\alpha (t)$ \ and using the inequality $\gamma \left| \dot{\alpha}\right|
\ll \omega _{0}^{2}\left| \alpha \right| $ one obtains (\ref{clas}).

\subsection{Operational regimes and measurable quantities}

The dynamics of doubly driven anharmonic oscillator exhibits a rich
phase-space structure, including regimes of regular and chaotic motion, with
two Rabi frequencies $\Omega _{1}$ and $\Omega _{2}$, and the difference $%
\delta $\ between driving frequencies and detuning $\Delta $ being the
control parameters. In study of the order-to-chaos transition in classical
systems a useful tool is the examination of a constant phase map in the
phase-space. In this context, our numerical analysis of the classical
equation of motion\ in the $(X,Y)$ plane ($X=%
%TCIMACRO{\func{Re}}%
%BeginExpansion
\mathop{\rm Re}%
%EndExpansion
\alpha $, $Y=%
%TCIMACRO{\func{Im}}%
%BeginExpansion
\mathop{\rm Im}%
%EndExpansion
\alpha )$ shows that the classical dynamics of the system is regular in
domains of small and large values of modulation frequency, i.e. $\delta \ll
\gamma $ and $\delta \gg \gamma $, and also when one of the perturbation
forces is much greater than the other: $\Omega _{1}\ll \Omega _{2}$ or $%
\Omega _{2}\ll \Omega _{1}.$ The dynamics is chaotic in the range of
parameters $\delta \gtrsim \gamma $ and\ $\Omega _{1}\simeq \Omega _{2}$,
where the classical results for $X=%
%TCIMACRO{\func{Re}}%
%BeginExpansion
\mathop{\rm Re}%
%EndExpansion
\alpha ,Y=%
%TCIMACRO{\func{Im}}%
%BeginExpansion
\mathop{\rm Im}%
%EndExpansion
\alpha $ obtained from Eq.(\ref{clas}) show that the classical strange
attractors for Poincar\'{e} section are realized \cite{OurPRE}. Thus, two
ways of producing chaos in a controlled manner can be considered by
monitoring the system through varying either the strength of the driving
force, or the difference frequency $\delta $.

We will examine macroscopic quantum effects contributing to the chaotic
behavior by considering both the Mandel $Q$ parameter \cite{Mandel} which
describes the deviation of excitation number uncertainty from the Poissonian
variance, i. e. $Q=\left( \left\langle \left( \Delta n\right)
^{2}\right\rangle -\left\langle n\right\rangle \right) /\left\langle
n\right\rangle ,$ $\left\langle \left( \Delta n\right) ^{2}\right\rangle
=\left\langle \left( a^{+}a\right) ^{2}\right\rangle -\left\langle
a^{+}a\right\rangle ^{2}$, and the Wigner function. Thus, we suggest that
the distinction between regular and chaotic dynamics can be most easily
understood by studying the dynamics of essentially quantum properties. In
fact, our numerical analysis has shown that the quantum dynamical
manifestation of chaotic behavior does not appear on ensemble averaged
oscillatory excitation numbers, but is clearly seen on the probability
distributions \cite{OurPRE}. The connection between quantum and classical
treatments of chaos can be realized by means of comparison between strange
attractors in the classical Poincar\'{e} section, and the contour plots of
the Wigner functions. Indeed, it was demonstrated in details \cite{OurPRE}
that for comparatively small values of the ratio $\chi /\gamma \lesssim 0.1$
the contour plots of Wigner functions are relatively close to the strange
attractors. Such likeness of quantum and classical distribution vanishes in
the deep quantum regime.

\section{Sub-Poissonian statistics for regular dynamics}

Our numerical analysis of quantum effects assisting{\bf \ }to chaotic
dynamics is based on QSD approach that represents the reduced density
operator averaging over the projectors onto the stochastic states $\left|
\Psi _{\xi }\right\rangle $ of the ensemble: $\rho (t)=M\left( \left| \Psi
_{\xi }\right\rangle \left\langle \Psi _{\xi }\right| \right) $, where $M$
denotes the ensemble averaging. The corresponding equation of motion is \cite
{Gisin}

\begin{eqnarray}
\left| d\Psi _{\xi }\right\rangle &=&-\frac{i}{\hbar }H\left| \Psi _{\xi
}\right\rangle dt  \label{QSD} \\
&&-\frac{1}{2}\sum_{i=1,2}\left( L_{i}^{+}L_{i}-2\left\langle
L_{i}^{+}\right\rangle L_{i}+\left\langle L_{i}\right\rangle \left\langle
L_{i}^{+}\right\rangle \right) \left| \Psi _{\xi }\right\rangle
dt+\sum_{i=1,2}\left( L_{i}-\left\langle L_{i}\right\rangle \right) \left|
\Psi _{\xi }\right\rangle d\xi _{i},  \nonumber
\end{eqnarray}
where $\xi $ indicates the dependence on the stochastic process, the complex
Wiener variables $d\xi _{i}$ satisfy the fundamental properties $M\left(
d\xi _{i}\right) =0,\;M\left( d\xi _{i}d\xi _{j}\right) =0,\;M\left( d\xi
_{i}d\xi _{j}^{\ast }\right) =\delta _{ij}dt,$ and the expectation value $%
\left\langle L_{i}\right\rangle =\left\langle \Psi _{\xi }\left|
L_{i}\right| \Psi _{\xi }\right\rangle $. For the oscillatory mean
excitations number and the variance this method gives:

\begin{equation}
\left\langle n\right\rangle =\lim_{m\rightarrow \infty }\frac{1}{m}\sum_{\xi
}^{m}\left\langle \Psi _{\xi }\left| a^{+}a\right| \Psi _{\xi }\right\rangle
,  \label{number}
\end{equation}

\begin{equation}
\left\langle \left( \Delta n\right) ^{2}\right\rangle =\lim_{m\rightarrow
\infty }\frac{1}{m}\sum_{\xi }^{m}\left( \left\langle \Psi _{\xi }\left|
(a^{+}a)^{2}\right| \Psi _{\xi }\right\rangle -\left| \left\langle \Psi
_{\xi }\left| (a^{+}a)\right| \Psi _{\xi }\right\rangle \right| ^{2}\right) ,
\label{variance}
\end{equation}

while the Wigner function is calculated with the formulas

\begin{equation}
W(r,\theta )=\sum_{m,n}\rho _{nm}W_{mn}(r,\theta ),  \label{wig}
\end{equation}
in terms of matrix elements $\rho _{nm}=M\left( \langle n\left| \Psi _{\xi
}\right\rangle \left\langle \Psi _{\xi }\right| m\rangle \right) $ of
density operator in the Fock state representation, where $(r,\theta )$ are
the polar coordinates in the complex phase-space plane $X=r\cos \theta $, $%
Y=r\sin \theta .$ The coefficients $W_{mn}(r,\theta )$ are Fourier transform
of matrix elements of the Wigner characterization function\thinspace \cite
{Garraway} 
\begin{equation}
W_{mn}(r,\theta )=%
%TCIMACRO{
%\QATOPD\{ \} {\frac{2}{\pi }(-1)^{n}\sqrt{\frac{n!}{m!}}e^{i(m-n)\theta }(2r)^{m-n}e^{-2r^{2}}L_{n}^{m-n}(4r^{2}),\;m\geq n}{\frac{2}{\pi }(-1)^{m}\sqrt{\frac{m!}{n!}}e^{i(m-n)\theta }(2r)^{n-m}e^{-2r^{2}}L_{m}^{n-m}(4r^{2}),\;n\geq m}}%
%BeginExpansion
{\frac{2}{\pi }(-1)^{n}\sqrt{\frac{n!}{m!}}e^{i(m-n)\theta }(2r)^{m-n}e^{-2r^{2}}L_{n}^{m-n}(4r^{2}),\;m\geq n \atopwithdelims\{\} \frac{2}{\pi }(-1)^{m}\sqrt{\frac{m!}{n!}}e^{i(m-n)\theta }(2r)^{n-m}e^{-2r^{2}}L_{m}^{n-m}(4r^{2}),\;n\geq m}%
%EndExpansion
,  \label{wcoef}
\end{equation}
where $L_{p}^{q}$ are the Laguerre polynomials.

\subsection{Controlled strong sub-Poissonian statistics\ \ \ \ \ \ \ \ \ \ \
\ \ }

First we discuss the case of classically regular behavior of the{\bf \ }%
considered model assuming the interaction of the system with vacuum
reservoir, $N=0$. In our numerical analysis\ we focus, for simplicity, on
the case of strong anharmonicity considering the parameters from $\chi
/\gamma =0.05$\ to $\chi /\gamma =0.7$. In the latter case the maximum mean
number of oscillatory excitations $\left\langle n\right\rangle =15$, while
in the former case it is equal to $\left\langle n\right\rangle =250.$

The results for time evolution of the mean oscillatory excitation number as
well as the Mandel $Q$ parameter averaged over quantum trajectories, are
depicted in Fig.1 and 2 for the parameters: (curve 1) $\chi /\gamma =0.7$, $%
\Delta /\gamma =-15$, $\Omega _{1}/\gamma =10.2$, $\Omega _{2}/\gamma =13.5$%
, and $\delta /\gamma =5$; (curve 2) $\chi /\gamma =0.05$, $\Delta /\gamma
=-15$, $\Omega _{1}/\gamma =38.18$, $\Omega _{2}/\gamma =49.5$, and $\delta
/\gamma =5$. As we see the $Q$ parameter shows a time-dependent modulation
with a period $2\pi /\delta $ and formation of sub-Poissonian statistics, ( $%
Q<0$) in which the oscillatory excitation number fluctuations are squeezed
below the coherent level $\left\langle \left( \Delta n\right)
^{2}\right\rangle <\left\langle n\right\rangle ),$ for time intervals
exceeding the transient time, $t\gamma \gtrsim 5.$\ The Mandel parameter
reaches its minimum $Q_{\min }\simeq -0.9$ and maximum $\ Q_{\max }\simeq
-0.38$ values for definite time intervals, for the case illustrated in Fig.
2 (curve 2). Moreover, the degree of sub-Poissonian statistics increases
with increasing of mean oscillatory excitation numbers at definite time
intervals. Below we demonstrate the formation of sub-Poissonian statistics
by analyzing the Wigner function. Figs. 3 show the Wigner function and its
contour-plot at the fixed moments of time $t_{k}=[7.13+(2\pi /\delta
)k]\gamma ^{-1},$ $(k=0,1,2,...)$ exceeding transient time. We find that the
Wigner function is located around the point $X=0,$ $Y=-10$ and its
contour-plot has narrow crescent form with the origin of phase space as its
centrum. Another peculiarity is that the Wigner function is nonstationary.
As calculations show, during the period of modulation $2\pi /\delta $ it
revolves{\small \ }around the origin of the phase-space. The radial
squeezing that represents the known property of the anharmonic oscillator to
produce the excitation number squeezing is also clearly seen in the figure.
The important novelty here is that the radial squeezing effect in this model
is much stronger, than an analogous one for the model of single driven
anharmonic oscillator. Indeed, for $\Omega _{2}=0$, when we turn to the
single driven anharmonic oscillator, the Mandel parameter is $Q=-0.72$ for
the same parameters as in Fig.2 (curve 2).

\subsection{Application to the generation of nonclassical light}

This model can be proposed for the realization of optical schemes generating
light with strong sub-Poissonian statistics. One of the candidates may be a
scheme involving self-phase modulation in an optical cavity pumped by two
coherent laser fields. In this scheme the anharmonicity of the oscillatory
mode comes from photon-photon interactions in the $\chi ^{(3)}$ medium
inserted in the cavity, while dissipative effects arise from leakage of
photons through the cavity mirrors, which damps the mode. In the case of a
ring cavity the evolution of the mode at frequency $\omega _{0}$ is
described by the Hamiltonian (\ref{hamiltonian}) with operators $a,$ $a^{+}$
being the annihilation and creation operators. The coupling constant $\chi $
of the Kerr interaction is proportional to the third-order nonlinear
susceptibility $\chi ^{(3)},$ and $\Omega _{1}$ and $\Omega _{2}$ are the
Rabi frequencies corresponding to two classical coherent fields at the
frequencies $\omega _{1}$ and $\omega _{2},$ respectively. The cavity
damping and noise effect are described by the Lindblad operator (\ref
{Lindblad}), where $\gamma $ is the damping rate and $N$ is the mean number
of thermal reservoir photons.

For the concrete scheme of light generation in a ring cavity, when the
coupling of ''in'' and ''out'' fields occurs at the different mirrors, the
cavity-output field operator $a_{out}=\sqrt{2\gamma }a$, while the
cavity-output intensity in photon number units per unit time is determined
by $n_{out}=\left\langle a_{out}^{+}a_{out}\right\rangle =2\gamma
\left\langle a^{+}a\right\rangle .$

It is easy to verify that in this case the results depicted on the Fig.1
relate also to the normalized intensity $n_{out}/2\gamma $ of output light.
For this scheme the measure of photocount statistics of detected output
photons can also be directly expressed in terms of itracavity photons,
namely Mandel factor $Q.$ Indeed, in sub-Poissonian light experiments the
observable quantity is the deviation of the variance of photocount number of
output photons defined as $Q_{i}=\left( \left\langle \left( \delta
n_{i}\right) ^{2}\right\rangle -\left\langle n_{i}\right\rangle \right)
/\left\langle n_{i}\right\rangle $, where $\left\langle n_{i}\right\rangle $
is the mean number of photocounts \cite{Mandel}. We consider a short
counting time $T\ll 2\pi /\delta $ to be accurate in detection of the
time-dependent output field modulated with period $2\pi /\delta $. In this
approximation $Q_{i}=\alpha T\,n_{out}\left( g_{out}^{(2)}-1\right) $, where 
$\alpha $ is a dimensionless quantum output of a detector, and $%
g_{out}^{(2)}=\left\langle a_{out}^{+}a_{out}^{+}a_{out}a_{out}\right\rangle
/n_{out}^{2}$ is the second order correlation function of the output field.
Further, in terms of intracavity photon numbers $Q_{i}=2\gamma \alpha TQ$,
and hence the normalized quantity $Q_{i}/2\gamma \alpha T$ reduces to the
Mandel parameter for intracavity photons. Thus, the results of Fig.2 are
applicable to quantum statistics of light and indicate the time-dependence
of the normalized quantity $Q_{i}/2\gamma \alpha T$. In particular, the most
strong sub-Poissonian statistics of generated light takes place for time
intervals at which the output intensity reaches its maximal values.

It is natural to explain such strong sub-Poissonian statistics of light by a
nontrivial interference effect or by the mutual interference of two coherent
fields in nonlinear medium. However such analysis is difficult to perform
because the system cannot be analyzed analytically in details. In this
context we restrict ourselves by the remark that the system under
consideration is an example of a quantum systems in contact with
environment, where the controlling and ordering of dissipative dynamics as
well as quantum statistics are realized through an external time-dependent
force. Some examples have been proposed in Refs. \cite{Controlled}.

\section{Nonclassical statistics for chaotic dynamics}

We now study the emergence of quantum chaos, which is expected to manifest
itself as crucial changes in the above results in the classically chaotic
operational regime. One of the ways to realize the controlling transition
from regular to chaotic dynamics is to vary the strength $\Omega _{2}$ of
the second force in the range from $\Omega _{2}\ll \Omega _{1}$ to $\Omega
_{2}\gg \Omega _{1}$. In the limit $\Omega _{2}\ll \Omega _{1}$ the system
is reduced effectively to the model of single driven anharmonic oscillator,
which exhibits bistability for a definite range of parameters $\ \chi
,\Delta ,\Omega _{1},$ and $\gamma $ \cite{Drummond}\ \ On increasing the
amplitude $\Omega _{2}$ the system transits from the regular to the chaotic
regimes at $\Omega _{1}\simeq \Omega _{2}.$ In the limit $\Omega _{2}>\Omega
_{1\text{ }}$the regular dynamics is restored. The results of numerical
simulation of the Mandel parameter in the regime of chaos are shown on
Figs.4 for the parameters: (curve 1) $\chi /\gamma =0.7$, $\Delta /\gamma
=-15$, $\Omega _{1}/\gamma =\Omega _{2}/\gamma =10.2$, and $\delta /\gamma =5
$; (curve 2) $\chi /\gamma =0.05$, $\Delta /\gamma =-15$, $\Omega
_{1}/\gamma =\Omega _{2}/\gamma =38.18$, and $\delta /\gamma =5.$
Surprisingly, the excitation-number fluctuations are also squeezed below the
coherent level for the considered chaotic regime. However, in contrast to
the previous regular regime, the excitation number exhibits both
sub-Poissonian ($Q<0$) and super-Poissonian ($Q>0$)\ statistics, that
alternate{\bf \ }in definite time intervals. The minimum and maximum values
of $Q$ in time intervals during one modulation period are equal to $Q_{\min
}\simeq -0.7$ and $Q_{\max }\simeq 0.6$ [for case (b)]. Thus, Figs.2 and 4
show the drastic difference between the behavior of Mandel parameter for
regular and chaotic dynamics. It means that the variance of oscillatory
number fluctuations may be used for testing of quantum chaos.

Now we illustrate the emergence of nonclassical sub-Poissonian statistics in
the doubly driven nonlinear oscillator in{\bf \ }its transition from regular
to chaotic dynamics using the phase-space symmetry properties of the Wigner
function. One can make sure of that by comparing the contour-plots of \
Wigner function for sub-Poissonian and super-Poissonian statistics. The
results of ensemble-averaged numerical calculations of \ both the Wigner
function and its contour-plot at fixed time intervals $t_{k}=[6.96+(2\pi
/\delta )k]\gamma ^{-1},$ $(k=0,1,2,...)$ are shown in Figs. 5 (a, b)
respectively. As we see the contour-plot for chaotic motion still has the
radial squeezed form [see Fig. 5 (b)]. This result takes place for $%
t_{k}=[6.96+(2\pi /\delta )k]\gamma ^{-1},$ $(k=0,1,2,...)$ at which the
Mandel parameter reaches its minimum value $Q_{\min }\simeq -0.7$. In
subsequent\ time intervals during the period of modulation the level of
excitation number fluctuations increases, and as a result the radial
squeezing in contour-plot decreases. This result is depicted in Figs. 6 (a)\
for the same parameters as in Fig.5(a), but for time-intervals $%
t_{k}=[7.325+(2\pi /\delta )k]\gamma ^{-1},$ $(k=0,1,2,...)$. The Poincar%
\'{e} section corresponding to these time moments are shown in Fig. 6 (b).

As we see there is a possibility to control the statistics of excitation
numbers of the oscillator by means of variation of the field strengths $%
\Omega _{2}$. One can obtain the following regime of generation: purely
sub-Poissonian statistics (Fig. 2 (curves 1, 2)), oscillating between super-
and sub-Poissonian (Fig. 4 (curves 1, 2)). Moreover, for certain values of
the parameters we achieve an essential improvement of sub-Poissonian
statistics in comparison with a single driven anharmonic oscillator, $\Omega
_{2}=0$. In this context it seems important to analyze in more detail the
Mandel parameter in order-to-chaos transition. To this end we have also
studied the behavior of the $Q$ factor versus the controlling parameter $%
\Omega _{2}/\gamma $. The results of numerical calculations at a definite
time moments exceeding the transient time, when the minimum of Mandel
parameter was realized, are displayed in Fig.7. As is seen the result is
displayed in the complicated form and shows the essential improvement of
sub-Poissonian statistics in the chaos-to-order transition taking place for
high values of $\Omega _{2}/\gamma $. The sub-Poissonian statistics goes bad
for small values of Rabi frequency $\Omega _{2}/\gamma $ as well as for its
intermediate values corresponding to the chaotic regime.

There is a particularly interesting feature to notice in the comparative
analysis of strange attractors and contour plots of the Wigner function
[Figs.5 (b,c) and Fig. 6 (a,b)]. We can conclude that it is possible to
predict the shape of Wigner function by the knowledge of Poincar\'{e}
section. Therefore, it is possible to predict the squeezing in quantum
system by the knowledge of Poincar\'{e} section of its chaotic classical
counterpart, that occurs by using Fig. 5 (c). This opens new possibilities
to guess intuitively what kind of states will appear in quantum chaotic
system by treating its classical chaotic map.

\subsection{Parameter scaling and dependence on external noise}

Further investigation of this model allows to establish other properties of
dissipative quantum chaos. It is easy to verify that Eq.(\ref{clas}) is
invariant with respect to the following scaling transformation of the
complex amplitude $\alpha \rightarrow \alpha ^{\prime }=\lambda \alpha ,$
where $\lambda $ is a real positive dimensionless coefficient, if the
parameters $\chi ,$ $\Delta $, $\Omega _{1},$ $\Omega _{2}$ are
correspondingly transformed as: $\chi \rightarrow \chi ^{\prime }=\chi
\diagup \lambda ^{2},$ $\Delta \rightarrow \Delta ^{\prime }=\Delta +\chi
\left( 1-1\diagup \lambda ^{2}\right) ,$ $\Omega _{1,2}\rightarrow \Omega
_{1,2}^{\prime }=\lambda \Omega _{1,2}.$\ As we see the scaling
transformations do not include the rate of the dissipative process $\gamma $
and hence can be written by using of dimensionless parameters as the
follows: $\Delta /\gamma \rightarrow \Delta ^{\prime }/\gamma ^{\prime
}=\Delta /\gamma +(\chi /\gamma )\left( 1-1\diagup \lambda ^{2}\right) $, $%
\chi /\gamma \rightarrow \chi ^{\prime }/\gamma ^{\prime }=\chi \diagup
\gamma \lambda ^{2}$, $\Omega _{1,2}/\gamma \rightarrow \Omega
_{1,2}^{\prime }/\gamma ^{\prime }=\lambda \Omega _{1,2}/\gamma $ .\ This
scaling property of the classical equation for the chaotic dynamics leads to
the symmetry of strange attractors which, for definite different sets of
parameters, have the same form in the phase-space and differ from each other
only in scale. Some examples of such behavior have been presented in our
previous paper \cite{OurPRE}. So, we conclude that the parameter scaling
happens for the classical trajectory dynamics for both regular and chaotic
regimes. It is interesting to analyze such scaling from the point of view of
quantum-statistical theory and its relevance for decoherence and
dissipation. It should be noted, that recently several studies have been
devoted to the scaling relationships in the decoherent quantum-classical
transition for chaotic systems (see \cite{Scaling-Chaos} and reference
therein). The quantum-classical transition is now understood to be affected
by the relative size of $\hbar $ for a given system as well as by the
parameter $D$, a measure of decoherence of quantum system of interest.
Further, in systems where the classical equation is chaotic, the transition
is also affected by the chaos in the system, and thus by $\theta $, the
Lyapunov exponent of the classical trajectory dynamics. Indeed, above
mentioned study \cite{Scaling-Chaos} was{\bf \ }devoted to scaling
relationships involving $\hbar ,$ $D,$ $\theta $. Here we hold on another
approach and present the results on scaling relationships, involving the
system parameters $\chi /\gamma $, $\Delta /\gamma $, and $\Omega
_{1,2}/\gamma $, in the framework of oscillatory numbers and its statistics.
The goal is to analyze what kind of parameter scaling we can achieve in a
quantum ensemble theory in the presence of quantum noise. We show below that
such parameter scaling occurs for wider ranges of the parameters, but for
not large values of the parameter $\chi /\gamma $. As our numerical
calculations show $Q$ Mandel factor is approximately invariant under scaling
transformations, i.e. $Q(\chi /\gamma ,\Delta /\gamma ,\Omega _{1,2}/\gamma
)\simeq Q(\chi /\lambda ^{2}\gamma ,\Delta /\gamma +\chi /\gamma
(1-1/\lambda ^{2}),\lambda \Omega _{1,2}/\gamma )$. This is illustrated in
Fig.8 (a) for a time interval during the period of modulation and for the
regular dynamics. Two sets of the parameters are: (curve 1) $\chi /\gamma
=0.1$, $\Delta /\gamma =-15$, $\Omega _{1}/\gamma =27,$ $\Omega _{2}/\gamma
=35$, and $\delta /\gamma =5$; (curve 2) $\chi /\gamma =0.05$, $\Delta
/\gamma =-15$, $\Omega _{1}/\gamma =38.18,$ $\Omega _{2}/\gamma =49.5$, and $%
\delta /\gamma =5$ with scaling parameter $\lambda =\sqrt{2}$. As we see,
the scaling invariance is valid within $0.02$. We also easily observe that
the degree of violation of the scaling increases with increasing of the
ratio $\chi /\gamma $ due to the moving of the system to a deeper quantum
regime. See, in particular Fig. 2 and Fig. 4 where we have shown the results
for two sets of parameters which are connected with a scaling transformation
with $\lambda =\sqrt{14}$. Analogous results take place for the ranges of
the parameters leading to chaotic dynamics. In Fig. 8(b) we illustrate the
scaling of the $Q$ factor for the parameters: (curve 1) $\chi /\gamma =0.1$, 
$\Delta /\gamma =-15$, $\Omega _{1}/\gamma =$ $\Omega _{2}/\gamma =27,$ and $%
\delta /\gamma =5$; (curve 2) $\chi /\gamma =0.05$, $\Delta /\gamma =-15$, $%
\Omega _{1}/\gamma =$ $\Omega _{2}/\gamma =38.18,$ and $\delta /\gamma =5$
with scaling parameter $\lambda =\sqrt{2}.$ The breaking of scaling
invariance in deep quantum chaotic regime is shown in Fig. 4 with scaling
parameter $\lambda =\sqrt{14}.$We emphasize that even though we have
motivated the study of the parameter scaling with connection to dissipation
and decoherence, the results obtained can be useful for other problems.

In the preceding sections we have studied statistics of excitation number in
the pure quantum regime assuming coupling of nonlinear oscillator with a
reservoir at zero temperature. In this part of the section we address the
question of how external noise affects statistics of excitation numbers. We
calculate the Mandel parameter with formulas (\ref{number}), (\ref{variance}%
) as in the previous case of vacuum reservoir. Calculating quantum
trajectories, we set the system initially to the vacuum oscillatory state
and integrate Eq. (\ref{QSD}) with Lindblad operators (\ref{Lindblad}) for
series of definite values of $N$. The results of numerical calculations at
definite time moments exceeding the transient time are depicted in Figs.9 ,
where minimum values of the $Q$ factor in time during the period $2\pi
/\delta $ versus external noise strength are presented.\ The curve (1)
describes the case of a single driven anharmonic oscillator ($\Omega _{2}=0$%
) for the same parameters as for the curve (2). As we see for the regular
regime (curve(2)) $Q$ increases linearly with $N$. The Mandel parameter is $%
Q=-0.88$ for $T=0$ K ($N=0$), but is 30 \% larger for $N=0.5$ (curve 2). An
analogous linear dependence takes place for the single driven anharmonic
oscillator \cite{Enzer}. However, for noisy chaotic regime the $Q$ factor is
approximately independent of noise and even slightly decreases for
comparatively large ranges of $N$ [Fig. 9 (curve 3)]. Such behavior can be
explained by additional noise-induced mechanisms in chaotic evolution,
namely, by chaos-to-order transition under influence of external noise.

\section{Conclusion}

Thus, in this work we have numerically studied the phenomena at the overlap
of chaos, dissipation and nonclassical statistics for the time-dependent
nonlinear model, namely Duffing oscillator driven by two periodic forces,
showing order-to-chaos transition. We have shown that physical systems based
on this model have a potential for generation of high-degree sub-Poissonian
light as well as for observation of quantum-statistical effects that
accompany chaotic dynamics. We have analyzed the mean oscillatory excitation
number, Mandel $Q$ factor and the Wigner function for both regular and
chaotic dynamics. We have shown in detail that the statistics of oscillatory
excitation number is drastically changed in the order-to-chaos transition.\
Particularly interesting is the essential improvement of sub-Poissonian
statistics in the regular operational regime of the scheme outlined above in
comparison with an analogous one for the standard model of a single driven
anharmonic oscillator. We have observed that in the chaotic regime the
system exhibits ranges of different statistics depending on the modulation
time, which alternate between sub- and super-Poissonian statistics. Another
important conclusion concerns the conformity between strange attractors and
contour plots of the Wigner function, that may be used for qualitative
testing of quantum states by treating its classical chaotic map. Considering
noise-induced effects we conclude that $Q$ parameter increases with
increasing of the noise level $N$ for regular dynamics. This result is in
good agreement with the results obtained for the single driven anharmonic
oscillator \cite{Enzer}. However, $Q$ parameter is almost insensitive to the
level of external noise for the case of chaos.

{\Large Acknowledgments}

G. Yu. K. gratefully acknowledges useful discussions with H.K. Avetisian and
support by the ISTC grant No. A-353. S. B. M. thanks Zh. Haroutunyan for
technical support.

\begin{center}
{\bf FIGURE CAPTIONS}
\end{center}

Fig. 1. Excitation numbers in regular regime for two sets of parameters (see
text). Excitation numbers for curve 1 is multiplied on 14. The averaging is
over 2000 trajectories.

Fig. 2. The Mandel factor for the same as in Fig.1 parameters. The averaging
is over 2000 trajectories.

Fig. 3. The Wigner function (a) and its contour-plot (b) for the regular
regime averaged over 3000 trajectories.

Fig. 4. The Mandel factor in chaotic regime for two sets of parameters (see
text).

Fig. 5. The Wigner function (a) and its contour-plot (b) averaged over 3000
trajectories. The Poincar\'{e} section (c) (approximately 20000 points)
plotted at times of the constant phase, when the maximal sub-Poissonian
statistics in quantum regime is reaslized. In both cases the dimensionless
parameters are in the range of chaos: 

Fig. 6. The contour-plot of Wigner function (a) and corresponding Poincar%
\'{e} section (b) (approximately 10000 points) constructed at time moments
when the maximal super-Poissonian statistics is reaslized. The parameters
and averagind for case (a) are the same as in Fig. 5.

Fig. 7. The behavior of the minimums of the Mandel factor $Q_{\min }$ versus
the controlling parameter $\Omega _{2}/\gamma $. The parameters are: $\chi
/\gamma =0.1,$ $\Delta /\gamma =-15,$ $\ \Omega _{1}/\gamma =27,$ $\delta
/\gamma =5.$

Fig. 8. The scaling of the $Q$ factor for two sets of parameters (see text)
in regular (a) and chaotic (b) regimes. The averaging is over 1000
trajectories.

Fig. 9. The dependence of $Q_{\min }$ on the noise intensity for: (curve 1)
single driven nonlinear oscillator ($\chi /\gamma =0.1$, $\Delta /\gamma =-15
$, $\Omega _{1}/\gamma =$ $27,$ and $\delta /\gamma =5$), (curve 2) doubly
driven nonlinear oscillator in regular ($\Omega _{2}/\gamma =35$), and
(curve 3) in chaotic ($\Omega _{2}/\gamma =$ $\Omega _{1}/\gamma =27$) 
regimes.

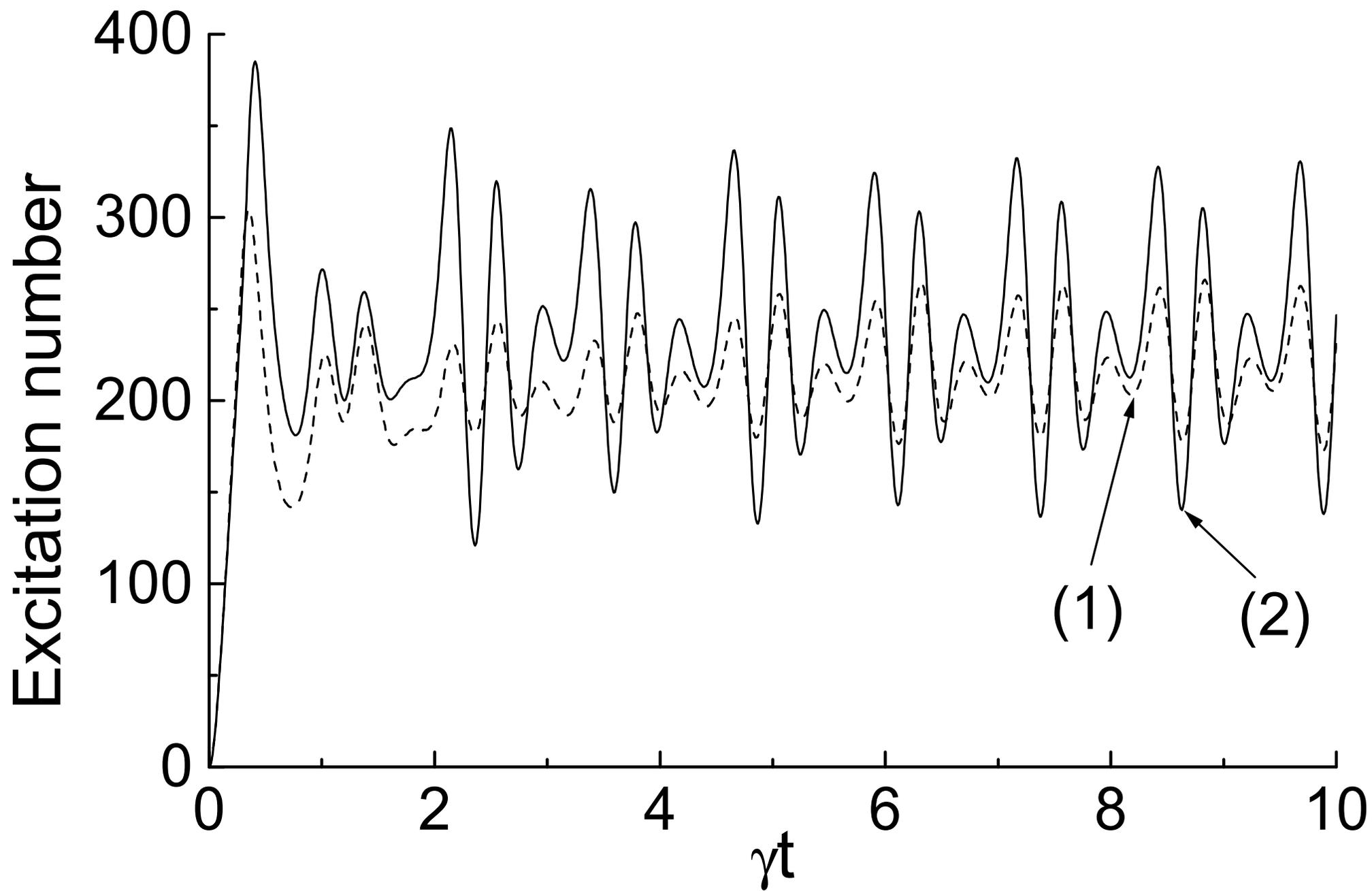
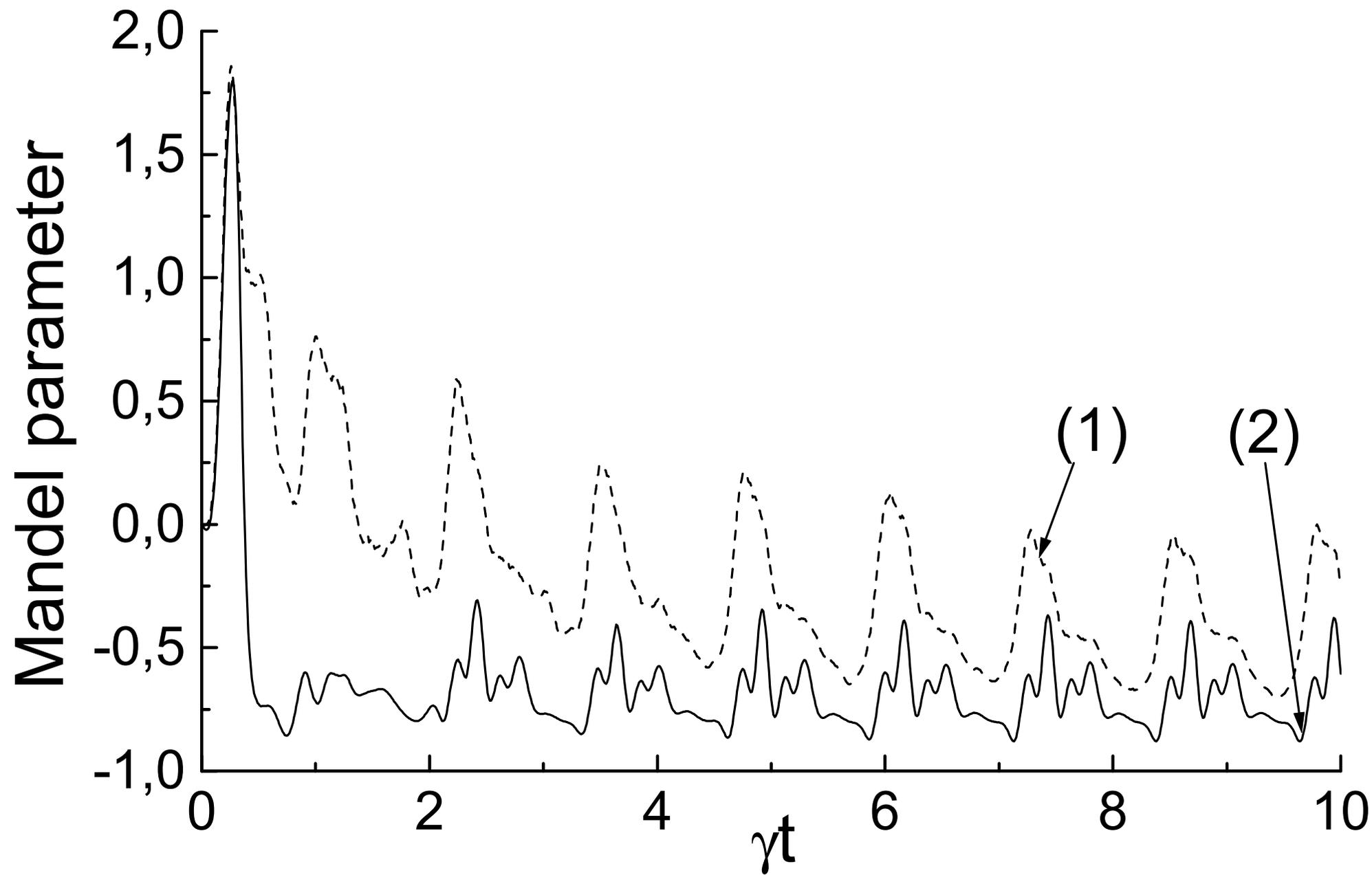
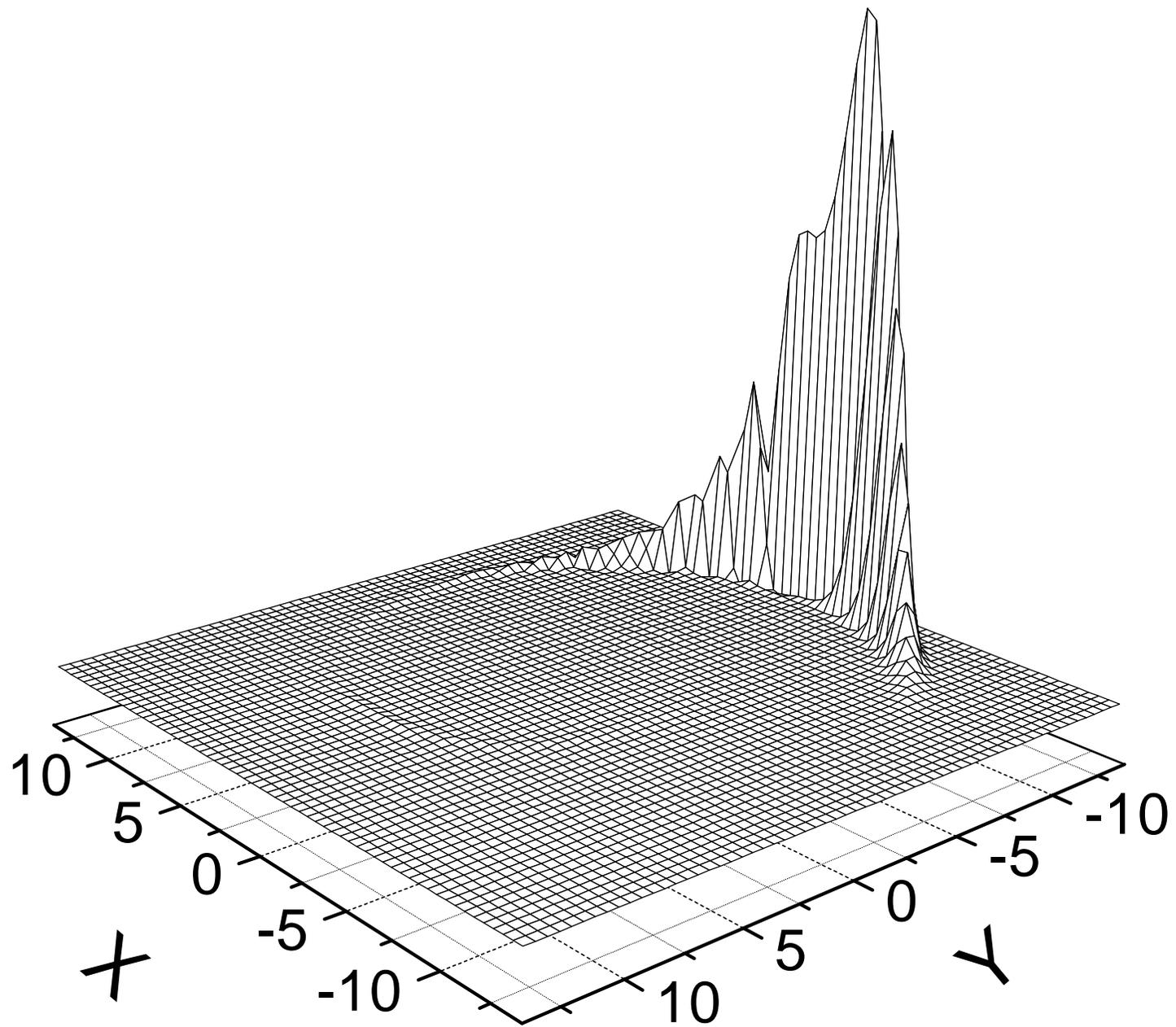
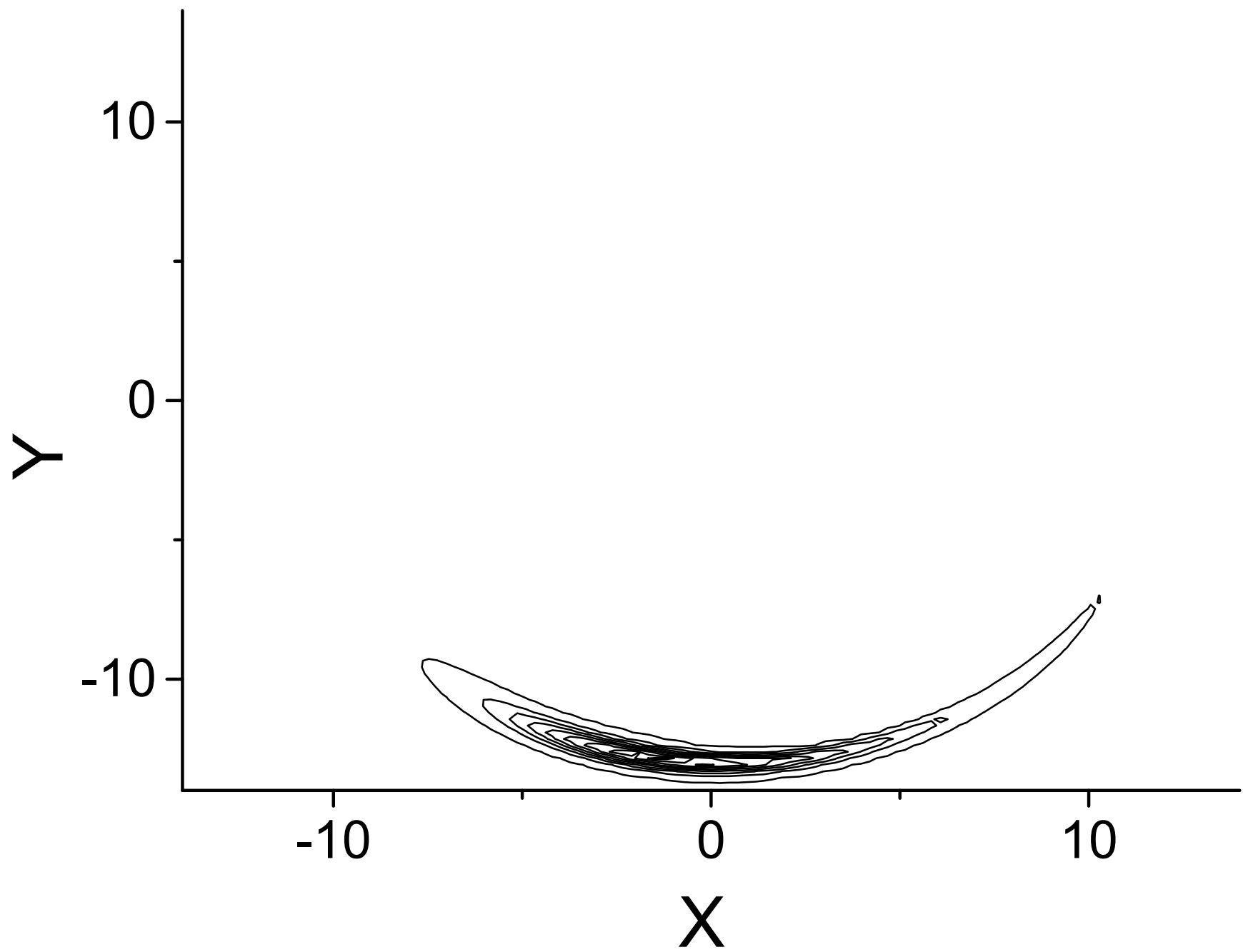
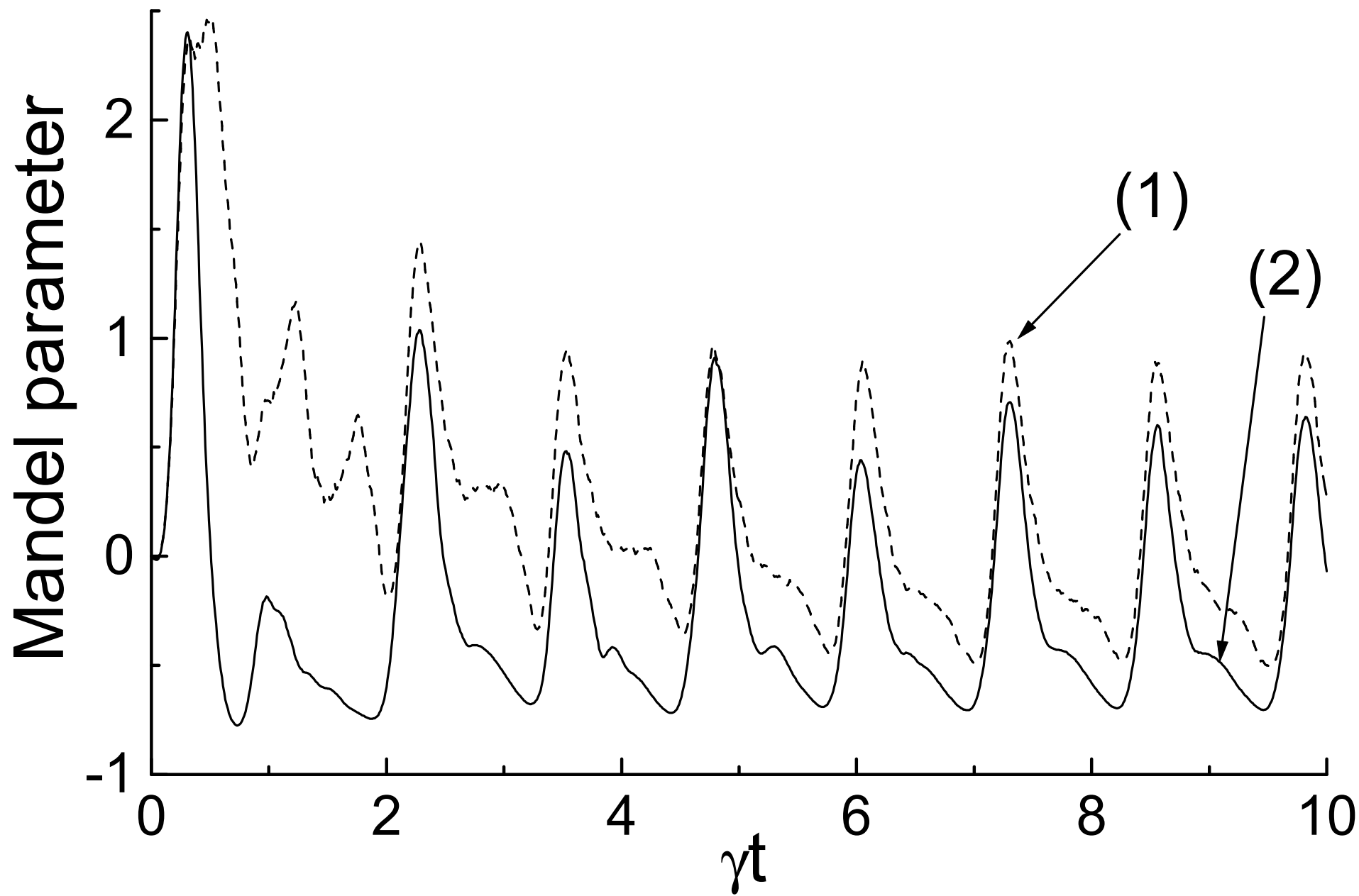
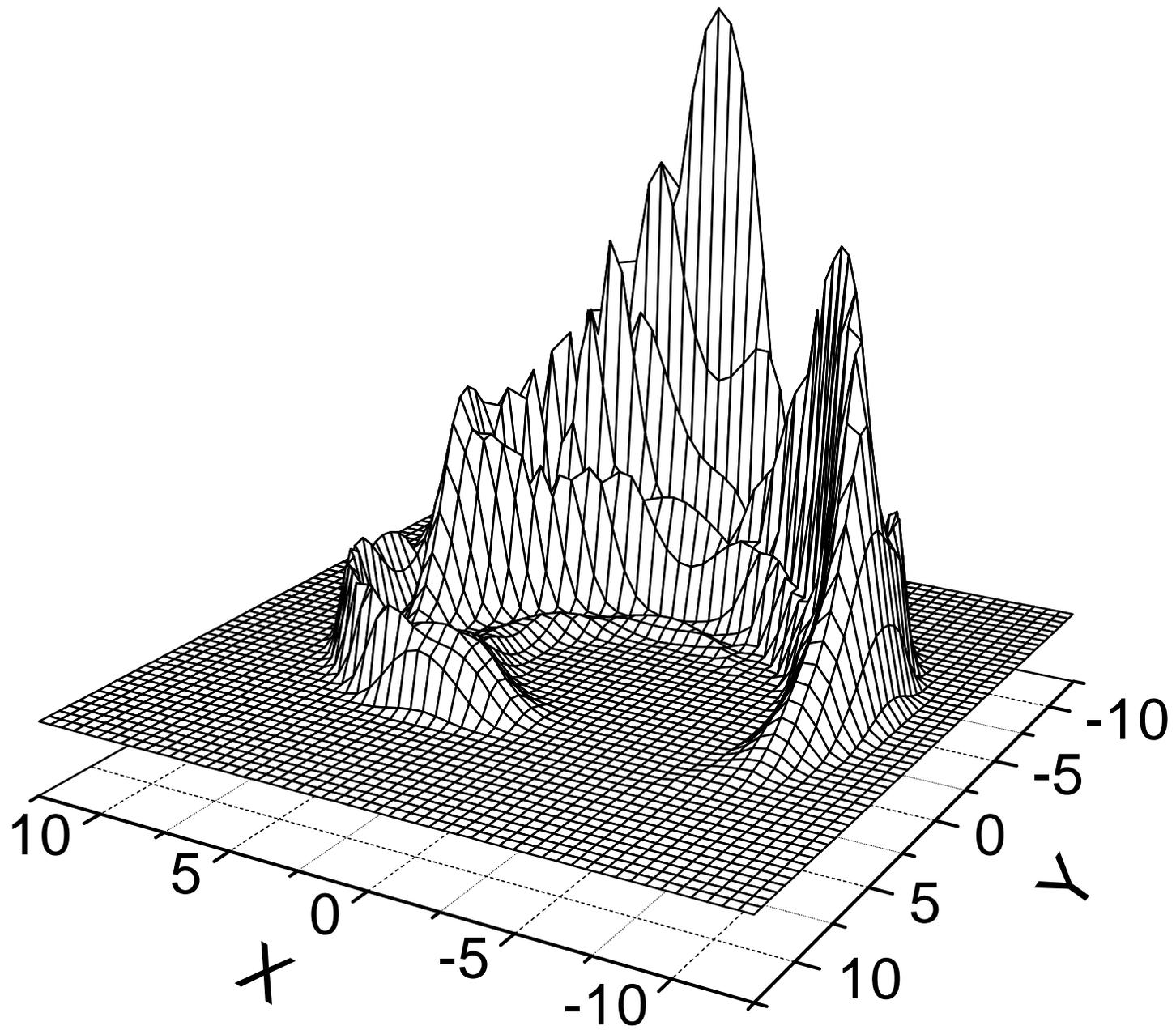
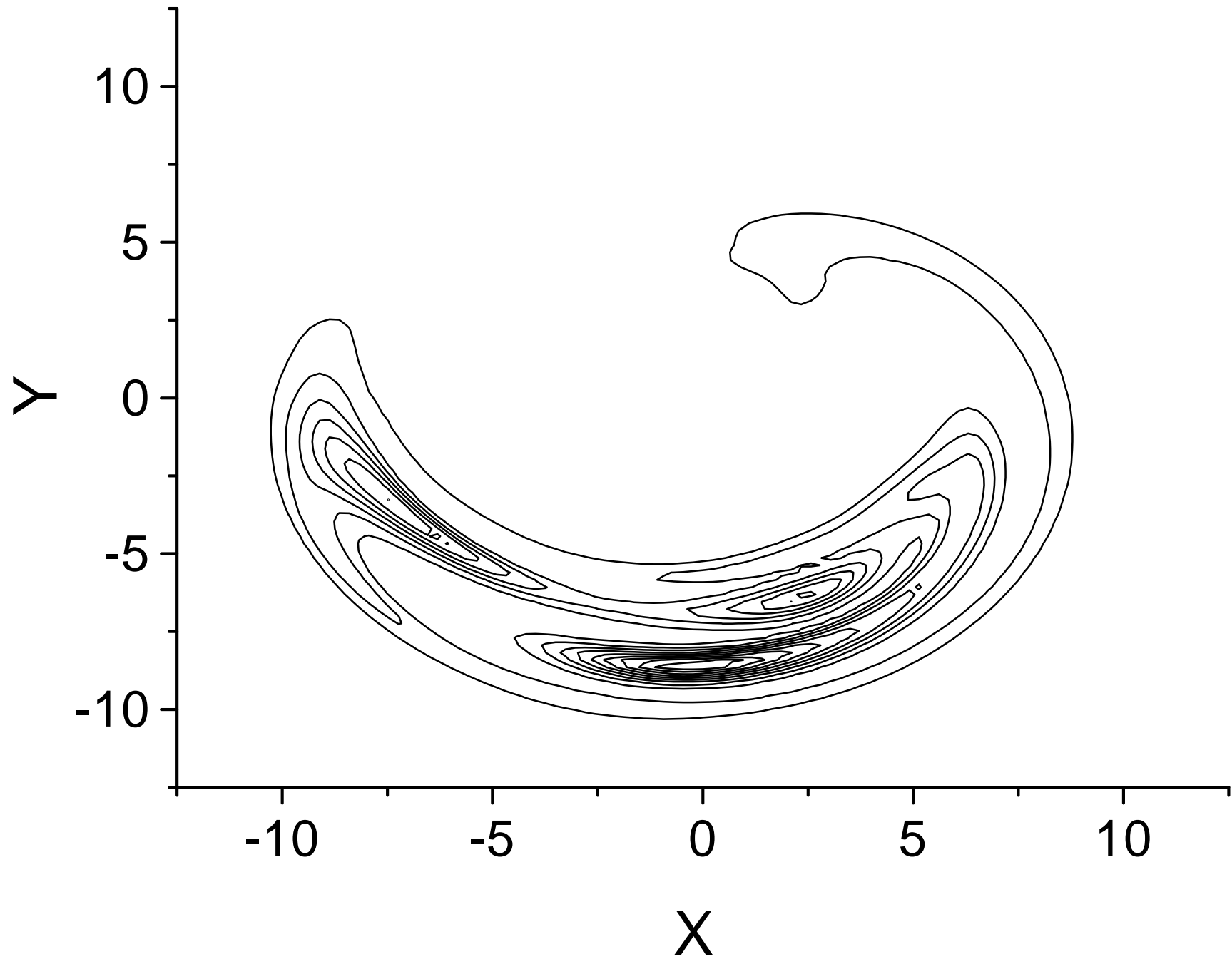
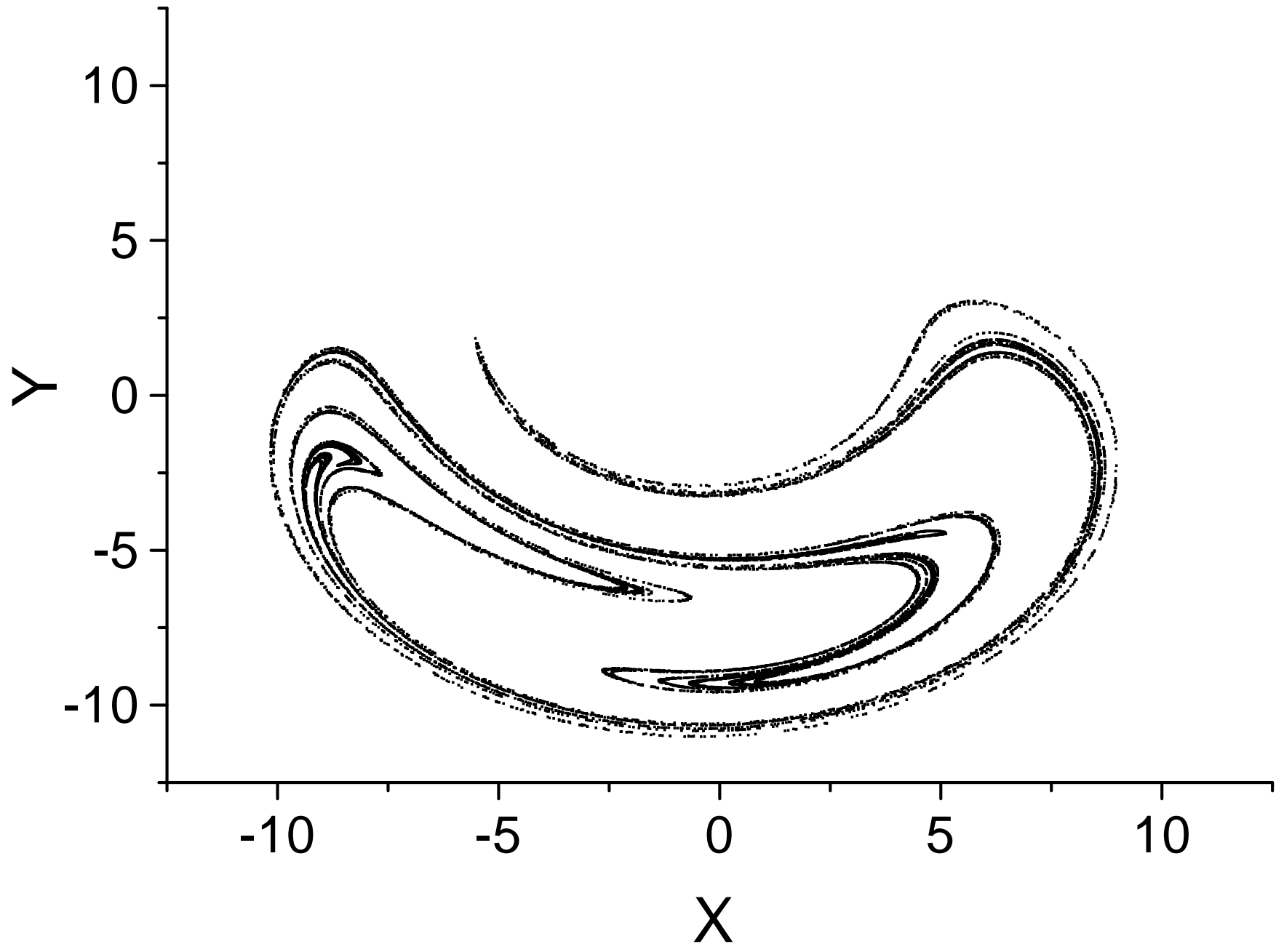
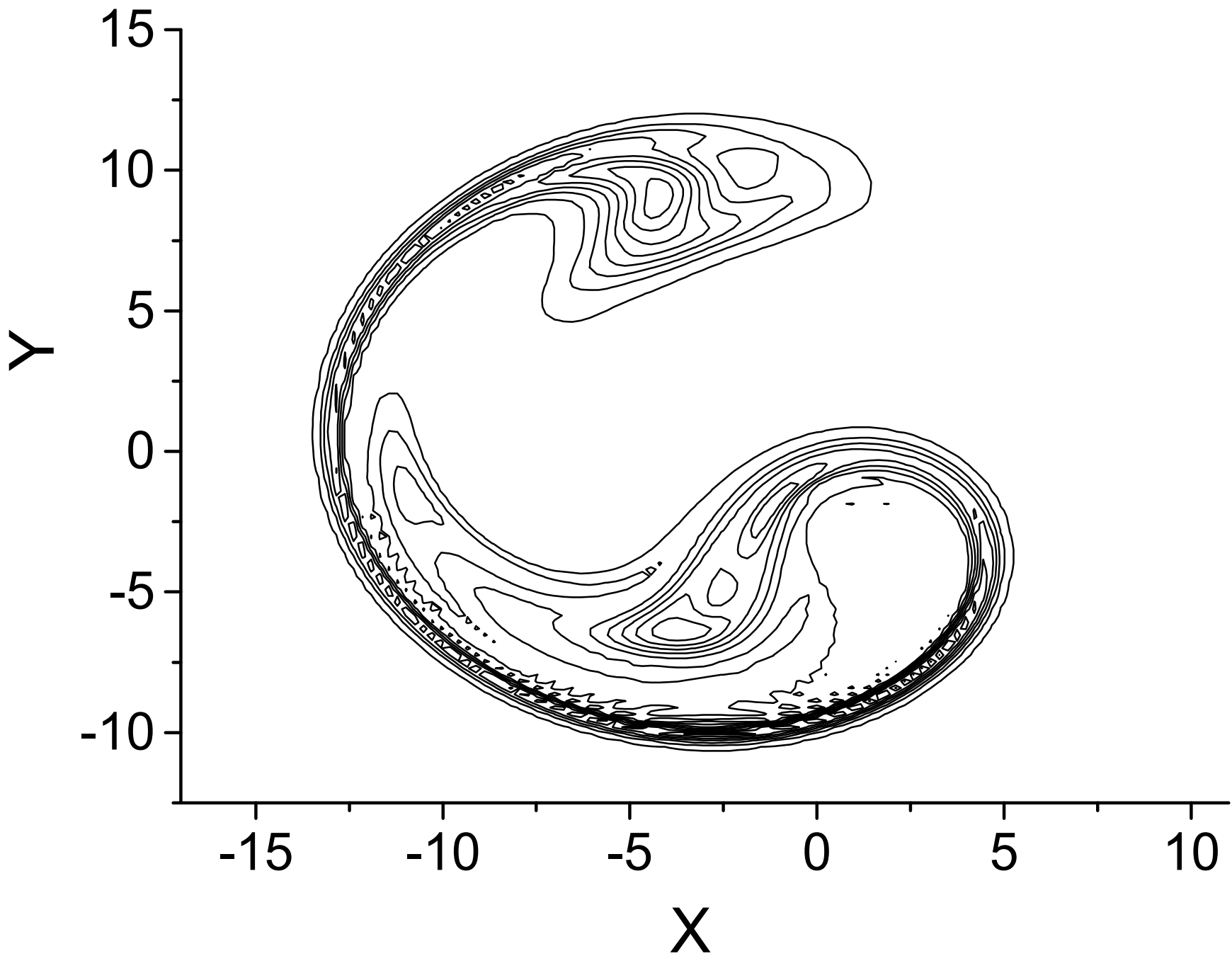
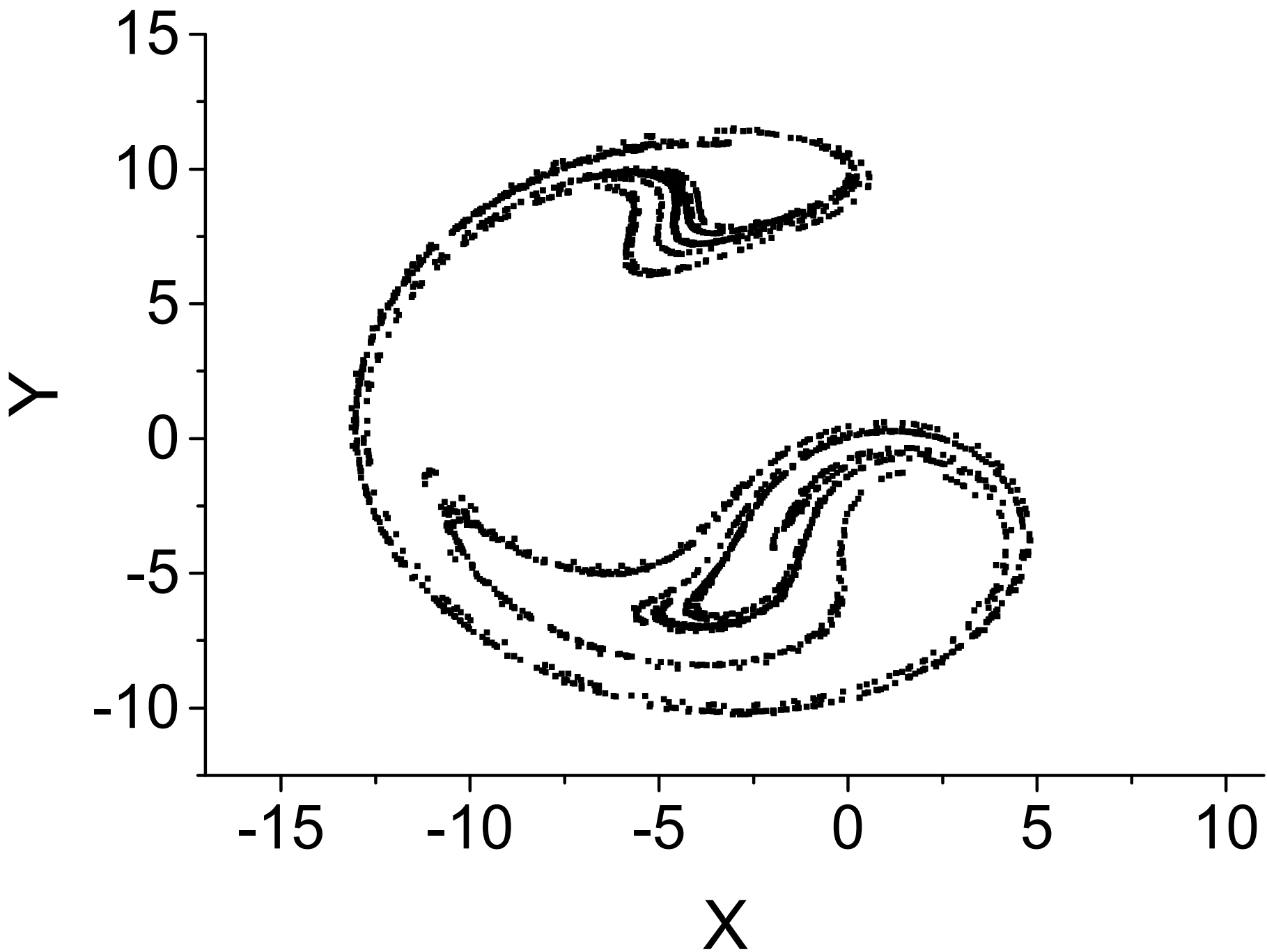
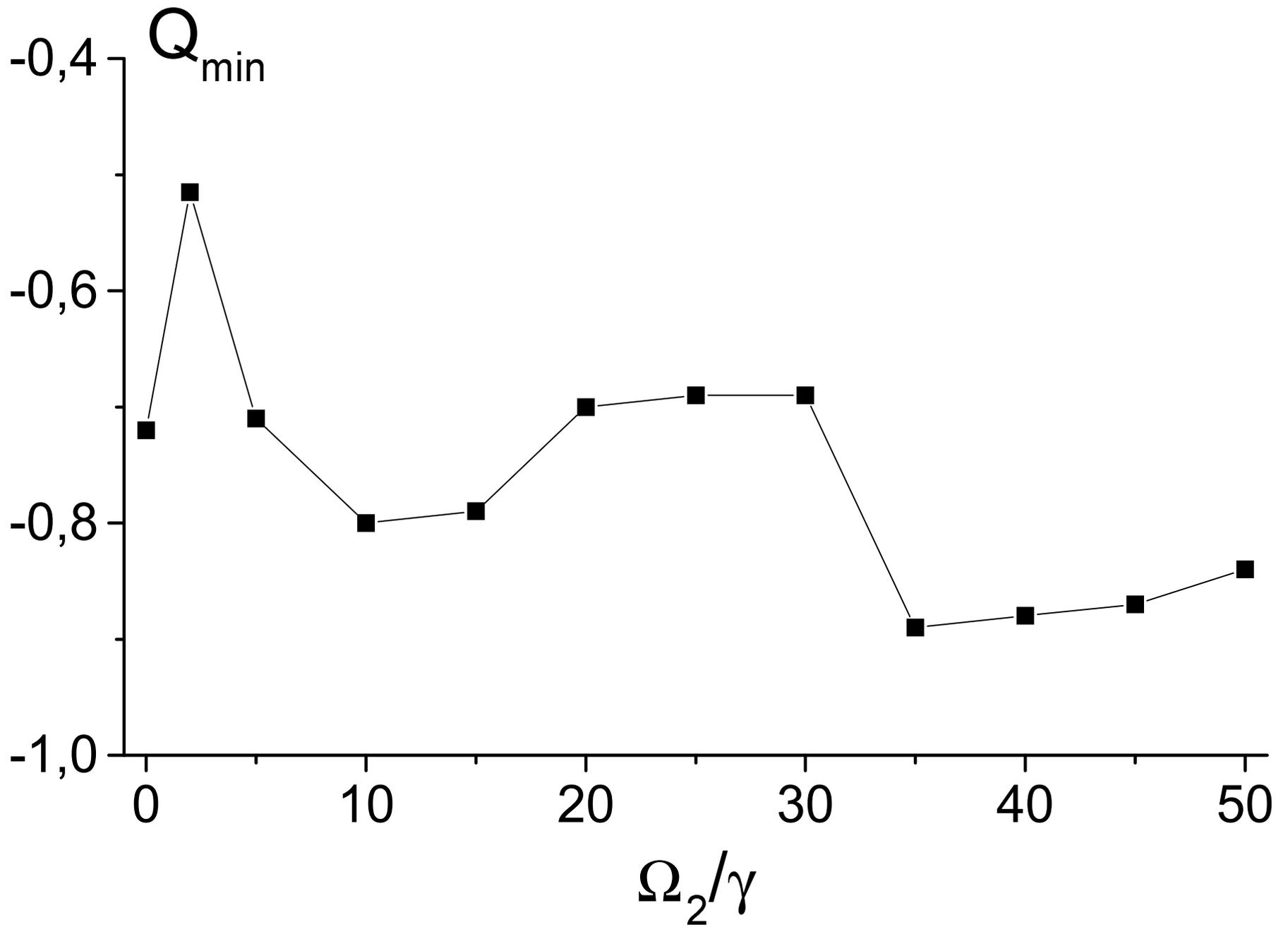
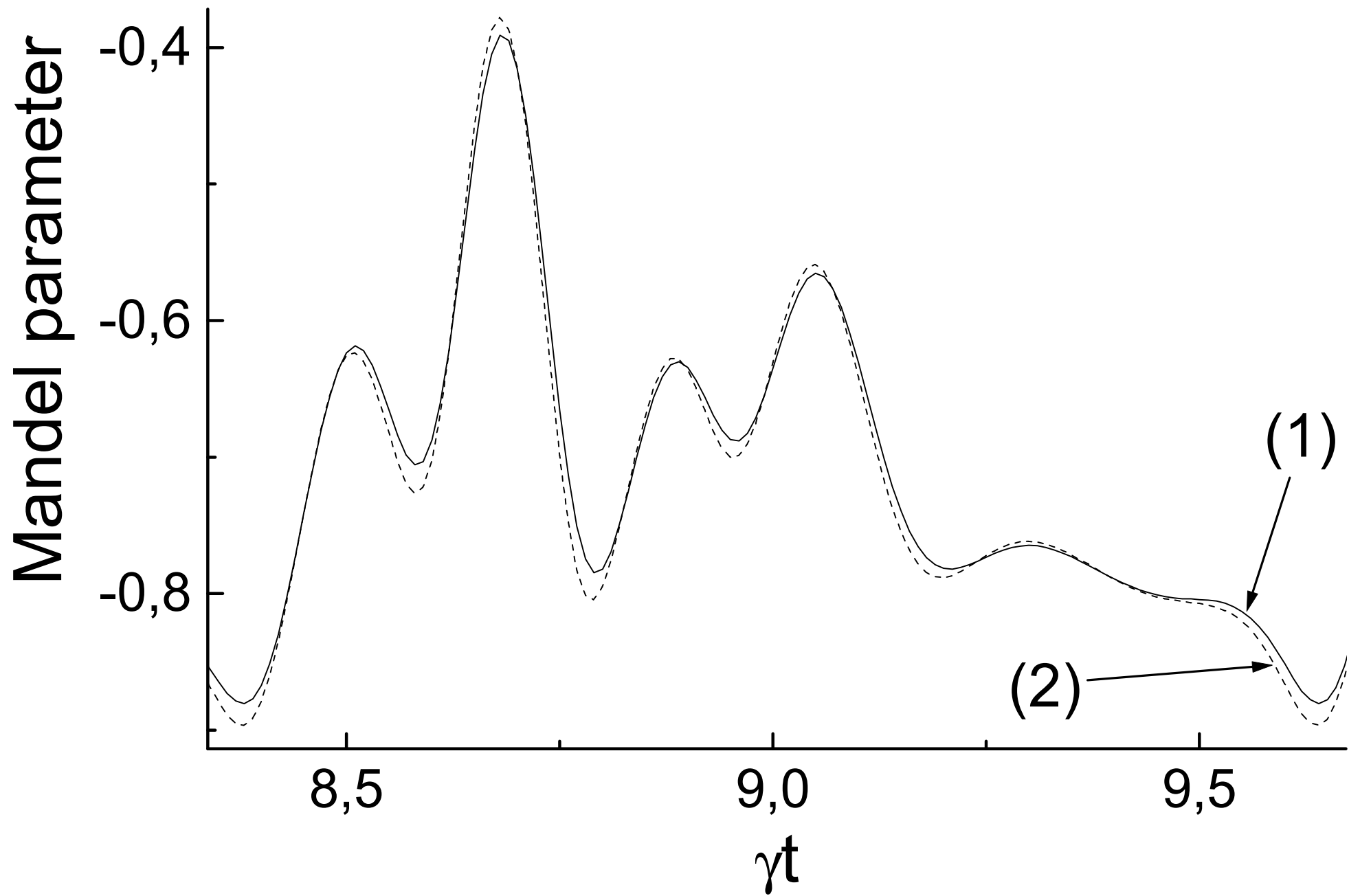
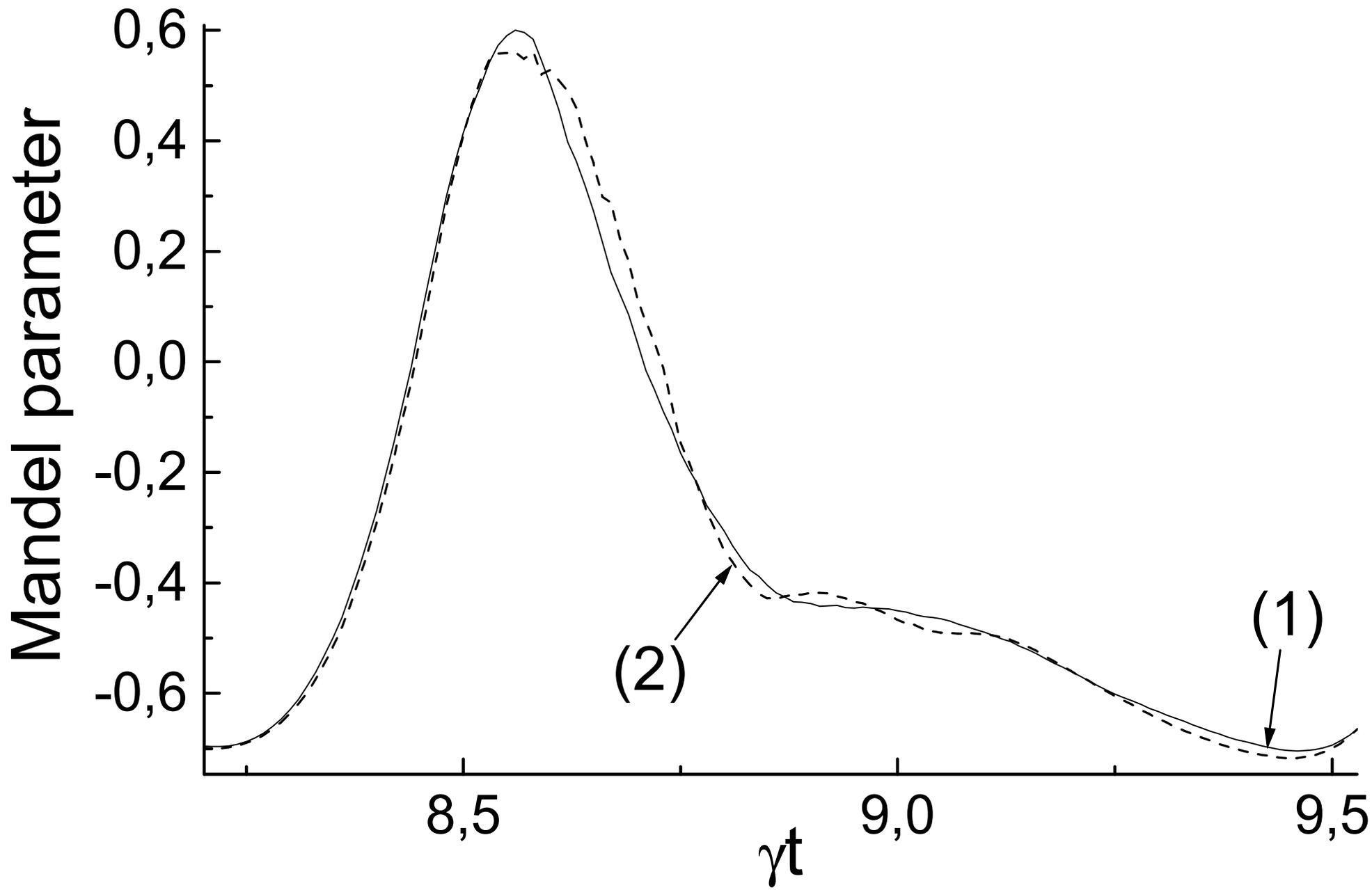
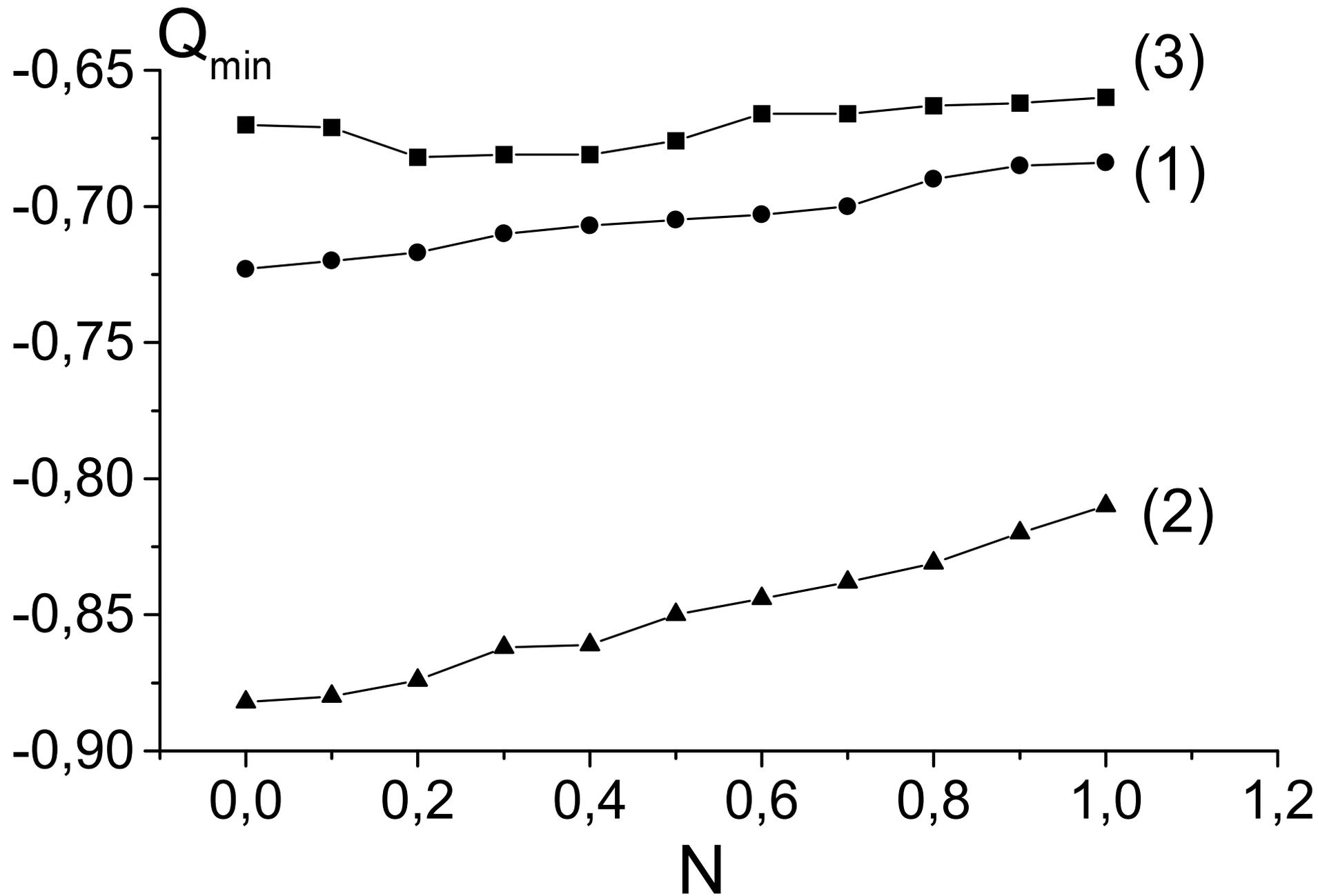
\end{document}